\documentclass[
12pt]{article}
\usepackage{epsfig}
\usepackage{graphicx}

\makeatletter

\@addtoreset{equation}{section}
\makeatother
\newcommand{\eq}{\begin{equation}}
\newcommand{\eqx}{\end{equation}}
\newcommand{\eqs}{\begin{equation*}}
\newcommand{\eqsx}{\end{equation*}}
\newcommand{\eqn}{\begin{eqnarray}}
\newcommand{\eqnx}{\end{eqnarray}}
\newcommand{\alg}{\begin{align}}
\newcommand{\algx}{\end{align}}

\newcommand{\f}[2]{\frac{#1}{#2}}

\begin{document}

\begin{titlepage}
\vskip1cm
\begin{flushright}
UOSTP {\tt 121201}
\end{flushright}
\vskip1.25cm
\centerline{\Large
\bf Entropy of universe as entanglement entropy}
\vskip1.25cm \centerline{ \large Dongsu Bak }
\vspace{1.25cm} \centerline{\sl  Physics Department,
University of Seoul, Seoul 130-743 Korea}

 \centerline{\tt dsbak@uos.ac.kr} \vspace{1.5cm}

\centerline{ABSTRACT} \vspace{0.75cm} \noindent
We note that the observable part of universe at a certain time $t_P$ is necessarily limited, when there is a beginning of universe. We argue that an appropriate spacetime region associated with an observer from $t_I$ to $t_P$ is the causal diamond which is the overlap of the past/future of the observer at $t_P$/$t_I$ respectively. We also note that the overlap surface $\partial D$ of the future and the past lightcones bisects the spatial section including  $\partial D$ into two regions $D$ and $\bar{D}$ where $D$ is the region inside the causal diamond and $\bar{D}$  the remaining part of the spatial section. We propose here that the entropy of universe associated with a causal diamond is given by an entanglement entropy where one is tracing over the Hilbert space associated with the region $\bar{D}$ which is not accessible by the observer. We test our proposal for various examples of cosmological spacetimes, including flat  or open 
 FRW universes, by showing
that the entropy as the area of $\partial D$ divided by $4G$ is a non-decreasing function of time $t_P$ as dictated by the generalized second law of thermodynamics.  The closed, recollapsing universe corresponds to a  finite system and there is no reason to expect the
validity of the generalized second law for such a finite system.

\vspace{1.75cm}
\end{titlepage}


\section{Our proposal}

In this note we would like to discuss the entropy of universe in a  cosmological context. In the usual discussion of the entropy of universe \cite{Fischler:1998st}-\cite{Egan:2009yy},  the precise specification of the system
is typically unclear.
Furthermore even if the system can be specified in a precise manner, the physical origin of the relevant entropy is often not  clear.

In this note we introduce a definition of the entropy of universe
as an entanglement entropy, based on the fact that the observable part of spacetime  at a given time $t_P$ is
necessarily limited, when there is a beginning of universe.
For the specification of the system in a cosmological context, we introduce the causal diamond \cite{Bousso:2000nf} at a present time $t_P$, which
 started at an initial time $t_I$ with $t_P > t_I$. The associated region of spacetime to an observer from $t_I$ to $t_P$ is given by the overlap of the past/future of the observer at $t_P$/$t_I$ respectively, which we shall call as the
causal diamond ${\cal D}(t_P,t_I)$. See Fig.~1a. In order to be idealistic, one may push $t_I$ into as far past as
possible, such that the initial time dependence is minimized. For instance, for the big-bang cosmology, $t_I$ might be taken as
the time of the big-bang which we set to be zero. For the case of de Sitter cosmology, the initial time might
be taken to be negative infinity as we will see below. In this note, we shall, however, keep the initial time
dependence $t_I$ to be general.

\begin{figure}[ht!]
\centering  
\includegraphics[width=10cm]{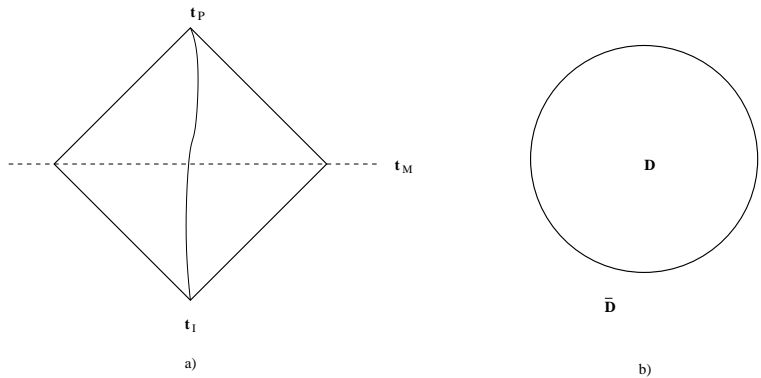}
\caption{\small The spacetime region for the small diamond ${\cal D}(t_P,t_I)$ is depicted on the left and, on the spatial section of the time $t_M$, the spatial region $D(t_M)$, which lies inside the small diamond,  is shown on the right. The surface $\partial D$
is the overlap of the past-lightcone from $O(t_P)$ and the future-lightcone from $O(t_I)$.}
\label{fig1}
\end{figure}

This small diamond is the spacetime region in which the observer can, in principle, directly probe all of its events starting from $t_I$.
The observation can be made for instance by sending out many observatory shuttles at $t_I$, which record all the events within
the diamond along their trajectories, and by recollecting these shuttles at the time $t_P$. The information of the region inside the
past of $O(t_P)$ and outside the future of $O(t_I)$ can be extracted, if one wishes, but one can have only partial information of this region,
which is already fully encoded by the events within the small causal diamond.
Thus this outside region is excluded for the sake of avoiding double counting.
Now note that the past lightcone emanated from $O(t_P)$ intersects with the future lightcone from $O(t_I)$ at some time $t_M$
assuming an appropriate time foliation of the spacetime. Then the small causal diamond ${\cal D}(t_P,t_I)$ bisects the spatial section
of the time $t_M$ into two regions.  As depicted in Fig.~1b,  the one inside  we shall call $D(t_M)$ whereas the outside region
will be denoted by $\bar{D}(t_M)$. We note also that the observer at $t_P$ cannot have any direct
physical information on the outside region $\bar{D}(t_M)$. Therefore the concept of  entanglement entropy can
be applied to this system on the spatial section at $t_M$. Since we do not have any information on
 $\bar{D}(t_M)$, we shall take the trace over the Hilbert space associated with  $\bar{D}(t_M)$
in order to obtain the reduced density matrix
\eq
\rho_D = {\rm tr}_{\bar{D}}\, \rho
\label{entangle}
\eqx
where $\rho(t_M)$ is the density matrix for the whole system on the spatial section at $t_M$.
Then we can define the entanglement entropy   in the standard way as
\eq
S_D = - {\rm tr}_{D}\, \rho_D \ln \rho_D
\eqx
As is well known, this entropy for the ground state of some (conformal) field theory is proportional to the area of the boundary of $D$ \cite{Srednicki:1993im}-\cite{Ryu:2006bv}:
\eq
S_D = \frac{1}{4G} \, A_{\partial D}
\label{area}
\eqx
where the Newton's constant $G$ depends on the system in a specific manner.
The entanglement entropy can in fact  include
the additional thermal-entropy contribution of the region $D$  if we take the original density matrix
as the finite-temperature density matrix instead of   the vacuum one.
For this finite-temperature 
\eq
S_D > \frac{1}{4G} \, A_{\partial D}
\eqx
due to the additional thermal contribution to the entropy.
With gravity included, the entropy including the thermal contribution is better described by
the area contribution alone \cite{Kabat:1994vj}. Namely for the system including dynamical gravity, which is for the present case, the areal entropy in (\ref{area}) is enough even for some system that is not in the ground state. For instance,
for the AdS black hole spacetime, the corresponding density matrix should be prepared in a specially entangled one \cite{Maldacena:2001kr},
which is not in a ground state, and the corresponding entropy is simply given by the area of the horizon divided by $4G$,
 which is the Bekenstein-Hawking entropy.

Even for the time dependent case, the areal entropy seems to represent the entanglement entropy
as discussed in Refs.~\cite{Bak:2007jm}.  But for the time dependent case, one needs perhaps a
better definition of entropy than the one in (\ref{entangle}) that is strictly valid only for the stationary
situation \cite{Bak:2007jm}.

\begin{figure}[ht!]
\centering  
\includegraphics[width=4cm]{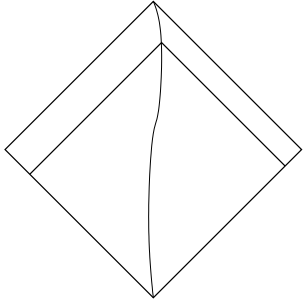}
\caption{\small The small diamond ${\cal D}(t_P,t_I)$ is completely included into the other diamond ${\cal D}(t'_P,t_I)$
if $t'_P > t_P$. }
\label{fig2}
\end{figure}

Thus having in mind these subtleties, we would like to claim that the entropy  $S_D$ that is basically coming
from the entanglement,  is the entropy observed at $t_P$ associated to the small diamond ${\cal D}(t_P,t_I)$,
and we propose that this is the entropy of universe at the time $t_P$ when the initial time $t_I$ is  pushed maximally
into the past.

 For $t'_P > t_P$, the small diamond ${\cal D}(t_P,t_I)$ is completely included into the diamond
 ${\cal D}(t'_P,t_I)$, which is the basis for the generalized second law of the thermodynamics.  As illustrated in
Fig.~2,  this implies in general that the time $t_M$ is non-decreasing  as a function of $t_P$, i.e. $t'_M \ge t_M$ if
 $t'_P > t_P$.
One natural question is then whether the corresponding entropy of the small diamond is satisfying the generalized
second law of the thermodynamics or not. In particular we shall test below whether the area of $\partial D(t_M)$ is
non-decreasing in time or not for various cosmological spacetimes. This way we shall show that our entropies of flat or open universes
are non-decreasing in time, which is in accordance with the generalized second law of thermodynamics.

\section{Test of the generalized second law of thermodynamics}

In this section, we ask weather our proposal is consistent with the generalized second law of the thermodynamics. For this we shall consider various cosmological spacetimes and test whether the area of  $\partial D (t_M)$ of the small diamond ${\cal D}(t_P,t_I)$ is non-decreasing or not. Note that the spatial extension of closed, recollapsing  universe is finite; One may wonder if the area of
$\partial D$ for this closed, recollapsing  universe may decrease in time,  especially in the
case that one can see more than half of the spatial section $S^{d-1}$. Below we shall argue that there is a possibility that  this can happen.
We shall also discuss the cases of de Sitter (dS) and anti-de Sitter (AdS) spacetimes and find that the results are also consistent
with the generalized second law of the thermodynamics.

We shall consider the spatially homogeneous and isotropic universe described by the Friedmann-Robertson-Walker (FRW) metric
\eq
ds^2= -dt^2 + a^2 (t) \f{dr^2}{1-k r^2}+a^2(t) d\Omega^2_{S^{d-1}}
\eqx
where $k=0,  -1,\ +1$ correspond to a flat, open or closed universe respectively. The Einstein equations become\footnote{We set $c=\hbar=1$
and, only for Section 3, the explicit dependence on $c$ and $\hbar$ will be recovered.}
\eqn
&&\f{\dot{a}^2 }{ a^2} + \f{k}{a^2}= \f{16\pi G}{d(d-1)} \rho\nonumber\\
&& \f{\ddot{a}}{a}= -\f{8\pi G}{d(d-1)} \Big(
(d-2)\rho + d \, p
\Big)
\label{einstein}
\eqnx
with the energy momentum conservation of matter
\eq
\f{d}{dt} \Big( \rho a^d \Big) + p \f{d}{dt} a^d =0
\eqx

Before presenting some detailed examples, we give a general identification of the time $t_M$ for the spatial section including $D(t_M)$
and the area of the boundary $\partial D(t_M)$. We here take the comoving observer located at $r=0$, and consider  a small diamond
${\cal D}(t_P,t_I)$ from an initial time $t_I$ to a  present time $t_P$. We shall take
\eq
t_0 \le t_I \le t_P \le t_\infty
\eqx
where $t_0$ is the time for the farthest past and $t_\infty$ for the farthest future.
The future lightcone beginning from $O(t_I)$ is then described by
\eq
\int^t_{t_I} \f{dx}{a(x)}= {\cal G}(r(t))
\eqx
and the past lightcone from $O(t_P)$ by
\eq
\int^{t_P}_{t} \f{dx}{a(x)}= {\cal G}(\tilde{r}(t))
\eqx
where ${\cal G}(r)\equiv\int^r_0 \f{dx}{\sqrt{1-k x^2}}$. Then by equating $r(t_M)=\tilde{r}(t_M)$ and by
introducing ${\cal B}(t)\equiv \int^t_{t_I} \f{dx}{a(x)}$, one finds
\eq
{\cal B}(t_M)= \int^{t_P}_{t_M} \f{dx}{a(x)} =\f{1}{2} {\cal B}(t_P)
\eqx
It is then clear that $t_M$ is non-decreasing as a function of $t_P$ as said before.
Due to the symmetry of the cosmological spacetime, the boundary $\partial D(t_M)$ has the shape of the $d-1$ sphere $S^{d-1}$
and its radius is given by
\eq
R(t_M)= a(t_M) r(t_M) = a\big[{\cal B}^{-1}\big(
{\mbox{\small $\f{1}{2}$}}
 {\cal B}(t_P)
\big)\big] \,\, {\cal G}^{-1}\big[
\mbox{\small $\f{1}{2}$} {\cal B}(t_P)
\big]
\eqx
Note that in the flat ($k=0$) case with  an energy condition $-|\rho| \le p \le |\rho|$,
the scale function $a(t)$ is always non-decreasing  in time and
$R(t_M)$ is non-decreasing as a function of $t_P$ since $t_M$, $a(t_M)$, and $r(t_M)$ are all
non-decreasing as  functions of $t_P$.
The corresponding entanglement entropy associated with the small diamond ${\cal D}(t_P,t_I)$  is given by
\eq
S_D= \f{A(t_M)}{4G} =\f{ s_{d-1} R^{d-1}(t_M)}{4G}
\eqx
where  $s_{d-1}=d\, \pi^{\f{d}{2}}/\Gamma(\f{d}{2}+1)$ is the area of unit $d-1$ sphere .

Now assuming a simple equation of state
\eq
p=w\rho
\eqx
with $w$ being a constant in time, one finds
\eq
\rho= \rho_0 \, \, \Big(\f{a_0}{a}\Big)^{d(1+w)}
\eqx

Let us first consider the flat universe with $k=0$.  The scale factor becomes
\eq
a= a_0 q_d\,\, t^{\f{2}{d(1+w)}}
\eqx
with
\eq
q_d = \left[ \f{4\pi G \rho_0 (1+w)^2 d}{d-1}\right]^{\f{1}{d(1+w)}}
\eqx
Since the scale factor $a(t)$ for $w < -1$ is decreasing from initially infinity to a finite size, we shall exclude this parameter
range for the subsequent discussion.

We now consider  a comoving observer located at $r=0$ and the corresponding small diamond from $t_I$ to $t_P$ with $t_I \ge 0$.
One finds that the time $t_M$ for the region  $D(t_M)$ is given by
\eq
t^{1-\f{2}{d(1+w)}}_M =\f{1}{2} \Big[ \,
t^{1-\f{2}{d(1+w)}}_P  + t^{1-\f{2}{d(1+w)}}_I
\Big]
\eqx
and the corresponding coordinate size of $\partial D(t_M)$ becomes
\eq
r(t_M)=\f{d(1+w)}{2 a_0 q_d \big(
d(1+w)-2
\big)} \Big[ \,
t^{1-\f{2}{d(1+w)}}_P  - t^{1-\f{2}{d(1+w)}}_I
\Big]
\eqx
We note that $\partial D(t_M)$ has the shape of $S_{d-1}$ and its radius $R(t_M)$ is given by
\eqn
&& R(t_M)=a(t_M)r(t_M)\nonumber\\
&&=\f{1}{
1-\f{2}{d(1+w)}}\f{1}{2^{\f{d(1+w)}{
d(1+w)-2}}} \Big[ \,
t^{1-\f{2}{d(1+w)}}_P  - t^{1-\f{2}{d(1+w)}}_I
\Big]\Big[ \,
t^{1-\f{2}{d(1+w)}}_P  + t^{1-\f{2}{d(1+w)}}_I
\Big]^{\f{2}{d(1+w)-2}}
\eqnx
It is straightforward to verify that this radius is monotonically increasing as a function of $t_P$.
For the choice $t_I=0$, one is led to
\eqn
R(t_M)= \left[
\begin{array}{cl}
\f{1}{
1-\f{2}{d(1+w)}}\f{1}{2^{\f{d(1+w)}{
d(1+w)-2}}} t_P &\quad\quad \ \ d(1+w) > 2 \\
0 &  \quad\quad 0 < d(1+w) < 2
\end{array}\right.
\eqnx
 The case with $d(1+w)=2$, leading to $R(t_M)=\infty$,  is excluded here since it requires a regularization.
Thus the entropy
\eq
S_D=\f{A(t_M)}{4G}=\f{s_{d-1} R^{d-1}(t_M)}{4G}
\eqx
is non-decreasing in time.

We now turn to the case of the flat inflationary patch of dS space\footnote{A similar interpretation
of dS entropy as the entanglement entropy across the dS horizon is put forward in Ref.~\cite{Conlon:2012tz}}.
This is governed by the Einstein equations (\ref{einstein})
with $w=-1$,  $k=0$ and $\rho=\rho_0 >0$.  The scale factor takes the form
\eq
a= a_0 \, e^{H t}
\eqx
with
\eq
H= \sqrt{\f{16\pi G}{d(d-1)}\rho_0}
\eqx
We take again the comoving observer located at $r=0$ and the small diamond ${\cal D}(t_P, t_I)$ of this
observer. The time $t_M$ for the spatial section including $D(t_M)$ is determined by the relation
\eq
e^{-Ht_M}=\f{1}{2} \Big(
e^{-H t_P} + e^{-H t_I}
\Big)
\eqx
and
\eq
r(t_M)= \f{1}{a_0 H} \Big(
e^{-H t_I} - e^{-H t_P}\Big)
\eqx
The boundary $\partial D(t_M)$ has the shape of the sphere $S_{d-1}$ with the radius $R(t_M)$
given by
\eq
R(t_M)= a(t_M) r(t_M)= \f{1}{H}\tanh \f{H}{2}(t_P-t_I)
\eqx
One can see that the radius is monotonically increasing as a function of $t_P$. Especially as $t_I$ goes to $-\infty$,
\eq
R(t_M) \ \rightarrow \  \f{1}{H}
\eqx
where the limit agrees with the radius of the dS horizon.
The entanglement entropy
\eq
S_D= \f{s_{d-1}R^{d-1}(t_M)}{4G}
\eqx
is then always increasing in time and bounded by the dS horizon entropy: 
\eq
S_D \le S_{dS}=\f{s_{d-1}}{4G H^{d-1}}
\eqx

Let us now turn to the case of a closed universe. The global patch of dS space is one example of a
closed universe, whose metric reads
\eq
ds^2= -dt^2 + \f{1}{H^2} \cosh^2 Ht \,\, \Big(
d\theta^2 + \sin^2 \theta \, ds^2_{S^{d-1}}
\Big)
\eqx
where $\theta \in [0, \pi]$. By introducing a new coordinate $\tau$ ($\in [-\f{\pi}{2},\f{\pi}{2}]$) defined by
\eq
\tau = 2 \tan^{-1} e^{Ht} -\f{\pi}{2}
\eqx
the above metric can be rewritten as
\eq
ds^2= \f{1}{H^2\cos^2 \tau} \Big[ -d\tau^2+
d\theta^2 + \sin^2 \theta \, ds^2_{S^{d-1}}
\Big]
\eqx
We consider an observer at $\theta=0$ and the diamond ${\cal D}(t_P,t_I)$ of this observer. We find that $\partial D(t_M)$ sphere has
the coordinates
\eq
\tau_M =\f{\tau_P+ \tau_I}{2}\,, \quad \quad \theta(\tau_M)=\f{\tau_P- \tau_I}{2}
\eqx
The  radius of the sphere $\partial D(t_M)$ is given by
\eq
R(t_M)= \f{1}{H} \, \f{\sin \f{\tau_P- \tau_I}{2}}{\cos \f{\tau_P+ \tau_I}{2}}= \f{1}{H} \Big(
\tan  \f{\tau_P+ \tau_I}{2} \cos \tau_I -\sin \tau_I
\Big) \quad \le \quad  \f{1}{H}
\eqx
which is monotonically increasing as a function of $\tau_P$. Since the trajectories of observers in the flat inflationary
and in the global patches
of dS space are chosen to be the same, the corresponding diamonds should be the same if the same initial/final points
of observation are chosen for both cases. Hence the above results for dS space are nothing but different
descriptions of the same physics by using different coordinate patches.

Let us now consider a closed, recollapsing universe which involves both big-bang and big-crunch singularities at the same time.
For this case, one can imagine a  universe which expands quickly but recollapses slowly, perhaps because
the equation of states changes around the time of the maximum expansion\footnote{We thank the referee for pointing out this possibility.}.
For such a universe, it is quite possible that the observer can see more than half of the spatial section $S^{d-1}$ at the time $t_M$ larger
than the moment of the maximum expansion. Then the generalized second law may be violated. But this closed, recollapsing
universe corresponds to a finite system and there is no reason to expect the validity of the generalized second law
for such a finite system. In the discussion section below, we shall present a more obvious example of such a finite case, where
one can see the violation of the generalized second law.

Finally we present one simple example with  open ($k=-1$)  universe.  For this we consider the cosmological patch of the
AdS space, whose metric is described by
\eq
ds^2 = l^2_{AdS} \left[ -dt^2 +  \cos^2 t \,\, \Big(
d\chi^2 + \sinh^2 \chi \, ds^2_{S^{d-1}}
\Big)\right]
\eqx
where $l_{AdS}$ is the AdS radius and $t \in [-\f{\pi}{2}, \f{\pi}{2}]$. We introduce a new coordinate $\tau$ by
\eq
ds^2 =\f{ l^2_{AdS}}{\cosh^2 \tau} \left[ -d\tau^2 +
d\chi^2 + \sinh^2 \chi \, ds^2_{S^{d-1}}
\right]
\eqx
with $\tau \in (-\infty,\infty)$.
With an observer at $\chi=0$, one has
\eq
\tau_M =\f{\tau_P+ \tau_I}{2}\,, \quad \quad \chi(\tau_M)=\f{\tau_P- \tau_I}{2}
\eqx
for $\partial D(\tau_M)$ sphere of the diamond ${\cal D}(\tau_P,\tau_I)$.
The radius of the $\partial D(\tau_M)$ sphere
is given by
\eq
R(\tau_M)= l_{AdS}\, \f{\sinh \f{\tau_P-\tau_I}{2}}{ \cosh \f{\tau_P+ \tau_I}{2}}
\eqx
which is increasing as a function of the time $\tau_P$.

\section{Entropy of our Universe}
In this section we would like to apply our definition of the entropy to our Universe to estimate the entropy we observe currently. Especially we would like to compare our estimation with the previous one based on the Bekenstein-Hawking entropy of the cosmological event horizon.
For this purpose we use the simple model
\eq
\dot{a}=\sqrt{\f{\Omega_\gamma}{a^2}+\f{\Omega_m}{a}+{\Omega_\Lambda}{a^2}}
\eqx
where we use the time unit given by the inverse of the Hubble constant $1/H$ and assume
  flatness ($k=0$) of the spatial section. Due to this flatness, our entropy is growing as a function of time
and the generalized second law of thermodynamics is respected. For the estimation, the cosmological parameters
are taken from Ref.~\cite{Seljak:2006bg}:
$h=H/(100km s^{-1}Mpc^{-1})=0.705\pm0.013$, $\Omega_m h^2=0.136\pm0.003$, $\Omega_\gamma$
can be computed from the blackbody radiation formula using the Cosmic Microwave Background (CMB) temperature $T=2.725\pm0.002 K$.
Then the vacuum energy density $\Omega_\Lambda$ is determined from
the flatness condition $\Omega_\Lambda=1-\Omega_m-\Omega_\gamma$.

To evaluate the entropy at the current moment, we consider the diamond with $t_I=0$ and $t_P =1/H$.
We find numerically that
\eq
t_M = 0.16 \pm 0.01 (H^{-1})\,,  \quad R(t_M)=0.38 \pm 0.02 (cH^{-1})
\eqx
For the comparison, the cosmological event horizon is evaluated as \cite{Egan:2009yy}
\eq
R_{CEH}(t=H^{-1})= c\, a(H^{-1})\int^\infty_{H^{-1}} \f{dx}{a(x)}=1.13\pm 0.05 (cH^{-1})
\eqx
leading to the entropy of the cosmological event horizon \cite{Egan:2009yy}
\eq
S_{CEH}(t=H^{-1})= 2.6\pm 0.3 \times 10^{122}
\eqx
while the entropy of Universe in our case is
\eq
S=2.9\pm0.3 \times 10^{121}
\eqx
Thus our estimation of entropy at the current moment is smaller than the corresponding
entropy of the cosmological event horizon by a factor about $10$.

\section{Discussions}
In this section we would like to discuss some general aspects of our proposal of the cosmological entropy.
For the flat FRW
spacetime, we have shown that our entropy for the small causal diamond of a comoving observer is
non-decreasing as a function of time, which is consistent with the generalized second law of thermodynamics.
For $k=0$ and $k=-1$, we gave various examples that are shown to be consistent with the generalized second
law of thermodynamics.  One may of course consider an observer who follows not just comoving
trajectory but a more general one. The causal ordering property of the small diamonds is intact.
Namely if $t'_P > t_P$, the diamond ${\cal D}(t_P,t_I)$ is fully included into the later diamond  ${\cal D}(t'_P,t_I)$ and consequently
$t'_M > t_M$.
Some general proof (or disproof) of the consistency with  the generalized second law of the thermodynamics
will be extremely interesting. Of course, we need to specify some necessary conditions for the generalized second law,
such as null energy condition for the matter part. One obvious example of spacetime that has to
be excluded is $R \times {T}^d$, where ${T}^d$ denotes the $d$ dimensional flat torus. Eventually
any observer in this spacetime can see the whole region of ${T}^d$, and the generalized second
law cannot work since the
entanglement entropy becomes zero afterward. (But there is nothing wrong with our proposal of the entanglement entropy
applied to this example since there is no reason to expect the validity of the generalized second law for such a
finite system in a finite box.)
A similar example of a finite system was discussed previously in the cosmological context of the closed, recollapsing universe.
Therefore we need some restriction on global properties of
spacetimes for the generalized second law. These require further studies.

Let us now turn to some aspects of
more general spacetimes. One interesting case is the spacetime including a black hole. Let us consider, for simplicity,
 a static black hole spacetime and an observer who remains outside the black hole horizon. Then the
causal diamond ${\cal D}(t_P,t_I)$ always remains  outside  the black hole,
 since the relevant past (future) lightcone cannot cross
the past (future) event horizon. For this outside observer, if $t_P-t_I \ll L/c$   where $L$ is the
observer's distance to the horizon, $\partial D(t_M)$ is not much affected by the presence of the black hole.  When
 $t_P-t_I \gg L/c$,   $\partial D(t_M)$ involves an extra inner boundary $\partial D_{BH}(t_M)$,
as depicted in Fig.~3. This inner boundary approaches more and more the black hole event horizon as
$t_P -t_I$ becomes larger and larger.

\begin{figure}[ht!]
\centering  
\includegraphics[width=8cm]{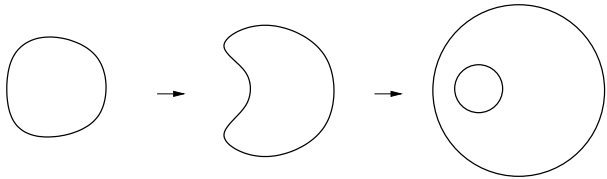}
\caption{\small
Considering the small diamond ${\cal D}(t_P,t_I)$ for an observer outside  horizon, the time development the region  $D(t_M)$  is depicted here and, for large enough time $t_P$, the inner boundary  of  $\ D(t_M)$ is completely  surrounding the black hole event horizon.}
\label{fig3}
\end{figure}

Thus one can see that our proposal includes the black hole entropy as an extra contribution as $t_P-t_I$ gets large enough.
The understanding of the Bekenstein-Hawking entropy of a black hole
as an entanglement entropy is rather well known. As was first noticed in Ref.~\cite{Israel:1976ur} and further developed in Refs.~\cite{Unruh:1976db,Jacobson:1995ak}, the entropy of the Schwarzschild black hole  may be understood as
an entanglement entropy between states inside and outside  the event horizon. More precise realization of this idea was given
in Ref.~\cite{Maldacena:2001kr} in the framework of the AdS/CFT correspondence  \cite{Maldacena:1997re}.  The large black hole in the AdS space involves two boundary
spacetimes in each of which a copy of a CFT lives. These two boundaries are causally
disconnected from each other due to the presence of the event horizon. Then the Bekenstein-Hawking entropy of the event horizon
is rather precisely realized by the entanglement entropy between the two boundary systems with an
initial state prepared in a particular manner \cite{Maldacena:2001kr}.
In Refs.~\cite{Bak:2007jm} a particular time dependent AdS black hole solution is  constructed where
$CFT_1$ and $CFT_2$ of the two boundaries are different from each other. One finds that the horizon area grows monotonically, which is due to  thermalization from an initially out-of-equilibrium state. One can again construct the reduced density matrix $\rho_1(t)$ by tracing over the Hilbert space of $CFT_2$. But now the von Neumann definition of the entanglement entropy
\eq
S_{ent}(t) \neq -{\rm tr}_1 \rho_1 (t) \ln \rho_1(t)
\eqx
is not applicable since, with this definition, the resulting entropy is necessarily time independent \cite{Bak:2007jm}. Rather the geometric entropy
 \eq
S_{ent}(t)  \sim \f{A_{h}(t)}{4G}
\eqx
is working better, which is similar to  our proposal.  Of course our proposal gives a refined version because there is a well defined prescription to identify the time $t_M$ of measuring $\partial D_{BH}(t_M)$ in relation with the observer's time  $t_P$  and  $t_I$.

Let us finally discuss the effect of boundary in the (asymptotically) AdS  geometry. Our prescription for this case is as follows. Whenever the causal diamond touches the boundary, we exclude the boundary region within the diamond from the definition of
$\partial D(t_M)$.  Based on this, we shall demonstrate that our entropy of small diamond for a particular choice of bulk observer is directly related to the holographic entanglement entropy in Refs.~\cite{Ryu:2006bv}.
To illustrate this explicitly, we consider the Poincare patch of $AdS_{d+1}$ whose metric takes the form
\eq
ds^2= \f{1}{z^2} \Big[ dz^2 -dt^2 + dr^2 + r^2 d\Omega_{S^{d-2}}\Big]
\eqx
where $z$ is the bulk coordinate ranged over $[0,\ \infty]$ and $z=0$ is the location of the boundary.
Now let us consider an observer following trajectory $z=z_0$ and $r=0$. The past lightcone from $t_P$ is described by
\eq
(t_P -t)^2= r^2 + (z-z_0)^2
\eqx
with $t_P \ge t$ and  the future lightcone from
$t_I$  by
\eq
(t -t_I)^2= r^2 + (z-z_0)^2
\eqx
with $t \ge t_I$. The time $t_M$ is identified as $t_M= \f{t_P+ t_I}{2}$ and the causal diamond is touching the
boundary if $T\equiv \f{t_P-t_I}{2} \ge z_0$. The region $\partial D(t_M)$ with the prescribed exclusion is described by
\eq
T^2 = r^2 + (z-z_0)^2
\eqx
with $z \ge 0$ or $z=\epsilon$ with regularization. Especially if one chooses an observer located at $z_0=0$ whom one may call as
boundary observer,
the region $\partial D(t_M)$ becomes
\eq
T^2 = r^2 + z^2
\eqx
with $z \ge \epsilon$.
It is then straightforward to prove that this corresponds to a minimal surface extended into bulk from the boundary $S_{d-2}$ sphere with radius $r=T$ that is excluded part from the boundary of $D(t_M)$.
Thus for this particular observer, the holographic entanglement entropy  \cite{Ryu:2006bv} is
\eq
S_{HE}= \f{A_{min}}{4G}
\eqx
where $A_{min}$ denotes the area of the minimal surface, and this agrees with ours
\eq
S_D=\f{A_{\partial D}}{4G}
\eqx
since two surfaces are identical. Of course for a generic observer instead of the boundary observer,
we do not have the interpretation of our entropy as the holographic entanglement entropy and then the above
cosmological interpretation is more appropriate.

\section*{Acknowledgement}
We would like to thank Andreas Gustavsson for careful reading of the manuscript.
This work was
supported in part by the 2012 UOS Research Fund.

\end{document}